\documentclass[12pt]{article}
\usepackage{amsmath}
\usepackage{amssymb}
\usepackage{graphicx}
\usepackage{natbib}
\usepackage{url} 
\usepackage[ruled, vlined]{algorithm2e}
\usepackage{float}

\newcommand{\blind}{0}

\begin{document}

\def\spacingset#1{\renewcommand{\baselinestretch}%
{#1}\small\normalsize} \spacingset{1}


\if0\blind
{
  \title{\bf scatteR: Generating instance space based on scagnostics}
  \author{Janith C. Wanniarachchi\hspace{.2cm}\\
    Department of Statistics, University of Sri Jayewardenepura, Sri Lanka\\
    and \\
    Thiyanga S. Talagala \\
    Department of Statistics, University of Sri Jayewardenepura, Sri Lanka}
  \maketitle
} \fi

\if1\blind
{
  \bigskip
  \bigskip
  \bigskip
  \begin{center}
    {\LARGE\bf scatteR: Generating instance space based on scagnostics}
\end{center}
  \medskip
} \fi

\bigskip
\begin{abstract}
Traditional synthetic data generation methods rely on model-based approaches that tune the parameters of a model rather than focusing on the structure of the data itself. In contrast, Scagnostics is an exploratory graphical method that captures the structure of bivariate data using graph-theoretic measures. This paper presents a novel data generation method, scatteR, that uses Scagnostics measurements to control the characteristics of the generated dataset. By using an iterative Generalized Simulated Annealing optimizer, scatteR finds the optimal arrangement of data points that minimizes the distance between current and target Scagnostics measurements. The results demonstrate that scatteR can generate 50 data points in under 30 seconds with an average Root Mean Squared Error of 0.05, making it a useful pedagogical tool for teaching statistical methods. Overall, scatteR provides an entry point for generating datasets based on the characteristics of instance space, rather than relying on model-based simulations.
\end{abstract}

\noindent%
\vfill

\newpage
\spacingset{1.5} 

\section{Graphical abstract}
\begin{figure}[htp]
    \centering
    \includegraphics[width=6in]{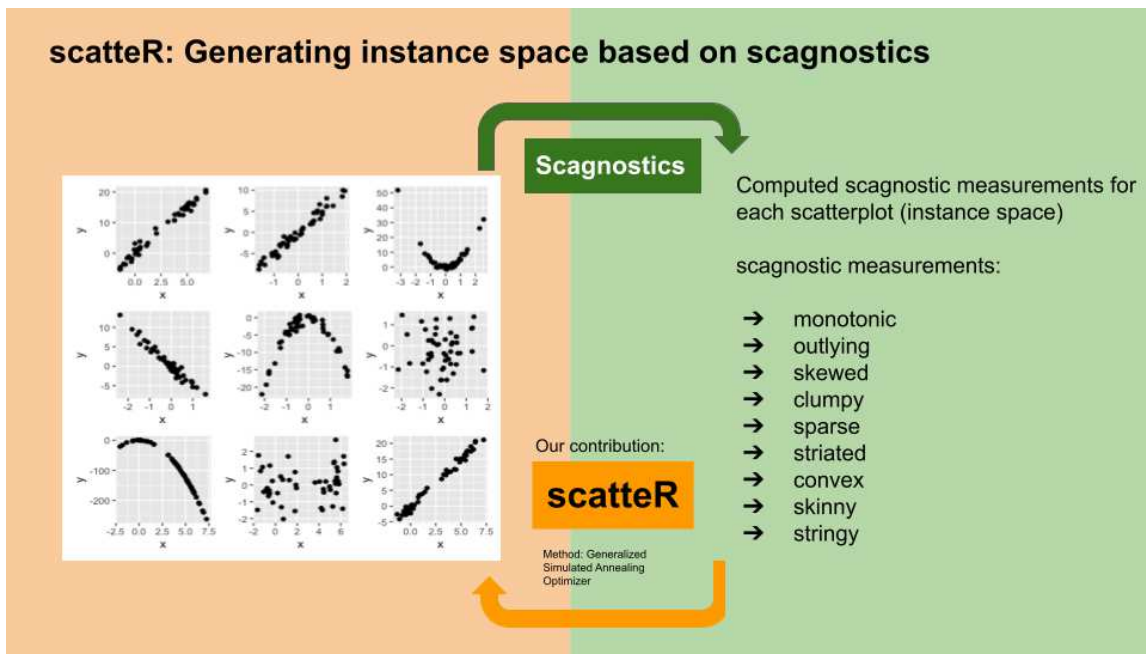}
    \caption{Graphical representation of this paper's contribution: In this study we introduce scatteR which is capable of generating instance spaces for a given set of scagnostics}
    \label{fig:graphical_abstract}
\end{figure}

\section{Introduction}
\label{sec:intro}
Scatter plots are a simple form of data visualization where two variables are plotted on a 2D plane as a set of points. The primary use of a scatterplot is to identify the insightful patterns and anomalies present within the two variables. In the presence of more than two variables, scatterplots of all possible combinations of two variables are combined to build a scatter plot matrix. As the number of variables is increased further, the task of identifying anomalies becomes visually taxing, which is the case in most modern-day datasets. Scagnostics (\cite{wilkinson2005graph}) was introduced as an exploratory graphical method to characterize the density, skewness, shape, outliers, and texture of a scatter plot through 9 measurements. Each scatterplot can be characterized by 9 scagnostic measurements, which provide information about their features. Therefore, the dimensionality of any scatterplot matrix is reduced to a fixed value (9) regardless of the number of variables in the original dataset. This brings forth the ability to identify anomalous scatter plots that exhibit a contrasting relationship with respect to the other relationships between variables. In practice, scagnostics have been used in big data settings with a large number of variables in exploratory data analysis and anomaly detection. (\cite{wilkinson2008scagnostics,lakshika2021computer}). 

Since scagnostics can quantify the characteristics of a scatter plot, an inverse scagnostics method would be ideal in a data synthesis process to generate data that resembles the characteristics of a scatterplot. 

In this study, we introduce scatteR, a novel method that is capable of providing a scatter plot that will give the desired values for a given set of scagnostic measurements. scatteR is available to be used as an R package (\url{https://www.github.com/janithwanni/scatteR})


The remainder of this paper is structured as follows. Section \ref{sec: litreview} discusses the landscape of data generation methods that are relevant to the study while Section \ref{sec:sig} explores the significance of this study. The methodology and framework of the scatteR algorithm are introduced in Section \ref{sec:algo} and the performance of the experiments are discussed in Section \ref{sec:results}. Finally, Section \ref{sec:conclusions} addresses the future directions and current known limitations of this method.

\section{Related Work}
\label{sec: litreview}



The suggested method would be centered at the intersection of the disciplines of visual analytics and optimization methods. A review of the literature surrounding this method will be discussed based on the theoretical explorations using scagnostics and the optimization algorithms that are available in this context. 

The modern graph-theoretic distributions were introduced by \cite{wilkinson2005graph} as an extension of the initial idea which was mentioned in \cite{tukey1985computer}. The introduction of scagnostics as a visual exploration tool brought forward many applications in preceding studies such as \cite{dang2012timeseer,sips2009selecting,srivastava2019ptr} even in recent years in analyzing features derived from images in a study done by \cite{lakshika2021computer}. The capabilities of scagnostics were fortified by \cite{wilkinson2008scagnostics} through the definition of a set of evaluation criteria for candidate scagnostic measurements. The data for the evaluation criteria has been developed using synthetic data derived from standard distributions and real-world datasets. The use of real-world datasets as a proxy for inverse scagnostic measurements has been used in many other instances such as \cite{dang2014scagexplorer,dang2014transforming,pham2020scagcnn} which greatly signifies the need for the method suggested in this paper as a proper methodology would have greatly benefited previous studies in their individual pursuits. As the popularity of the scagnostic measurements grew complementary methods were introduced that sought to resolve the issues that were discovered within scagnostics. \cite{wang2019improving} discovered through the use of bivariate normal distributions that two of the scagnostic measurements Outlying and Clumpy are largely sensitive to the binning process within scagnostics as the measurements changed drastically when the underlying data was met with small perturbations. Similarly, the sensitivity of scagnostics towards scale transformations was explored in \cite{dang2014transforming} where the best transformations that provided the most information from scagnostics were discovered. An implementation of scagnostics in Javascript by \cite{pham2020scagnosticsjs} extended the idea of scagnostics to higher dimensions. Modern deep learning has been used in conjunction with scagnostics in a novel method introduced by \cite{pham2020scagcnn} where better estimates of scagnostic scores are given based on convolutional neural networks \cite{simonyan2014very}. This study is the closest study that mentioned the use of synthetic data to match scagnostic measurements however the method of scagnostic measurements was not mentioned in the study.

\section{Significance}
\label{sec:sig}
Currently in the R programming language (\cite{rlangcite}) there are two known functional implementations of scagnostics. The package that is considered in this study (\cite{scagnosticscite}) is implemented using a Java (\cite{arnold2005java}) backend. Moreover in the Python (\cite{van1995python}) programming language there is an implementation of scagnostics as well. In addition, there is also an implementation in Javascript which contains visualizations that extend into higher dimensions (\cite{pham2020scagnosticsjs}). However, even though there are implementations of the scagnostics in R, Python, and even Javascript, an implementation of a package that would be able to generate scatterplots given a set of scagnostic measurements is currently not available. If such a method exists, then this method could be used for pedagogical reasons, such as demonstrating statistical methods in different instance spaces within a classroom. In addition, researchers and educators would be capable of generating the dataset that is required to demonstrate their downstream tasks and to verify the robustness of their techniques. However, there is still a potential gap, up to now, in solutions to synthesize a dataset that would give the desired value for a specific set of scagnostic measurements. Through the scatteR package, it would be possible to expedite the testing of new statistical methods on different structures of instance spaces while, on the other hand, helping extend the knowledge of statisticians to understand the effectiveness of statistical methods in different instance spaces.

\section{Algorithm}
\label{sec:algo}

\subsection{Scagnostics}
Scagnostics were first introduced by \cite{tukey1985computer} as an exploratory graphical method that is capable of giving a quantifiable value for the different features of a scatterplot. Later \cite{wilkinson2005graph} improved this idea by developing the nine scagnostic measurements based on graph-theoretic measures. There are nine scagnostic measures - Outlying, Skewed, Clumpy, Sparse, Striated, Convex, Skinny, Stringy and Monotonic. These features are calculated based on the three geometric graphs of a Convex hull, $\alpha$ shape, and a minimum spanning tree. A geometric graph is a representation of a graph in a numerical space where vertices refer to points in the coordinate plane and edges refer to the lines connecting a pair of points. The measurements are calculated using equations based on features such as the length between edges, the perimeter, area, diameter, and length of a graph. For example, the measurement of Convex is calculated using the following equation
\[
\text{Convex} = \frac{area(A)}{area(H)}
\]
where $area(A)$ is the area of the $\alpha$ shape and $area(H)$ is the area of the Convex hull. Similarly,  the Skinny measurement is calculated using the following equation,
\[
\text{Skinny} = \frac{1 - \sqrt{4\pi area(A)}}{perimeter(A)}
\]
From the above equations and the other equations mentioned in \cite{wilkinson2005graph}, it is evident that generating scatterplots that would give exact measurements is nontrivial and is heavily constrained.

\subsection{Problem Statement}
Scagnostics is a function $s(\mathbf{D}): \mathbb{R}^{n\times2}\rightarrow\mathbb{R}^9$ where $\mathbf{D} \in \mathbb{R}^{n\times2}$ matrix with each row representing an $(x,y)$ coordinate. The goal would be to then create an inverse function $s^{-1}(m): \mathbb{R}^k\rightarrow\mathbb{R}^{n\times2}$ such that given an expected measurement $m_0$, $s(s^{-1}(m_0)) \approx m_0$ where $m,m_0\in \mathbb{R}^k$ and $k \in \{1,2,\dots,9\}$. 

\subsection{Search based approach}
The given problem statement could be naively modeled as a search problem where the possible states are instances of $\mathbf{D}$ (i.e. scatterplots) and the possible pathways from one state to another is defined by moving a single $(x,y)$ coordinate in any of the 8 directions (upper right, up, upper left, right, left, lower right, down, lower left, stay). However, this poses a problem as the number of possible actions grows as the sample size increases for there are $n^8$ possible paths from a single state where $n$ is the sample size. 

\subsection{Optimization Approach}
Therefore, the inverse function could be modeled as an optimization problem, where we would be needing to arrange a set of $n$ points in a unit 2D plane such that the loss (Section \ref{sec:loss}) is minimized.

The parameters to be optimized would then be a vector of coordinates presented as $[x_1,x_2,\dots,x_n,y_1,y_2,\dots,y_n]$ thereby ensuring that each coordinate is optimized individually. In this case, a gradient-free optimization method has to be taken due to the nature of the calculations performed to obtain the scagnostic measurements. In such a configuration the number of parameters then grows linearly as the sample size increases which requires optimization methods that are capable of exploring a large range of a high dimensional space in a limited amount of time. 

Thus, a smaller version of the inverse function can be modeled based on an iterative refinement method. In this method, an initial number of $n_0$ points are arranged in the unit 2D plane such that the loss is minimized for the dataset $\mathbf{D_1}$ given by the $n_0$ points. Afterward, each subsequent iteration will concatenate the previous datasets and arrange the $n_0$ number of points that minimizes the loss using the current dataset comprising of previous datasets and the $n_0$ number of points. This will result in a series of datasets $[\mathbf{D_1},\mathbf{D_2},\dots,\mathbf{D_M}]$ where $M = [N / n_0]$ and $N$ is the total number of observations that are needed in the final dataset. To increase the computational efficiency there is also the possibility of giving the starting solution for the next iteration based on the result of the previous iteration with a small error added to ensure that the optimization algorithm does not get stuck in a local optimum.

In terms of optimization methods, several branches could be explored: gradient-based methods, gradient-free methods, and metaheuristic methods (\cite{koziel2011computational}). The ideal optimization algorithm for this study would have to be centered around the gradient-free methods and metaheuristic methods as the formulation of scagnostics does not produce differentiable functions. There are several known limitations of derivative-free methods as mentioned in \cite{conn2009introduction} which includes difficulty in optimizing non-convex functions. When faced with non-convex functions derivative-free methods tend to deviate towards a local optimum in their initial stages and from thereon would try to smooth the function thereby stagnating in a local optimum. Additionally, it is not reasonable to optimize a high dimensional function using derivative-free methods due to the computational effort that is needed. In such instances, metaheuristics are ideal as they provide a near optimal answer within the given constraints. 

Metaheuristics are high-level search methods that explore the search space of an optimization problem using different strategies. When choosing these strategies, an ideal balance should be kept between exploration (also known as diversification) and exploitation (also known as intensification), thereby ensuring that the regions with high quality solutions are found quickly and that the regions with low quality solutions aren't exploited unnecessarily. 

\subsubsection{Simulated Annealing}
Many optimization problems contain objective functions that have many local minima separated by several barrier heights. Simulated annealing was brought forward to address this issue to find a global minimum in a complex multidimensional objective function. The term annealing is derived from annealing in metallurgy, where certain properties of a material are controlled through changes in the temperature of the material. Likewise, the simulated annealing algorithm works by setting an initial temperature value and gradually decreasing the temperature towards zero. As the temperature is decreased, the algorithm becomes greedier in selecting the optimal solution. In each time step, the algorithm selects a solution closer to the current solution and would select the new solution based on the quality of the solution and the temperature-dependent acceptance probabilities. This allows the algorithm to select worse solutions at the beginning of the search, thereby ensuring that a larger solution space is explored before exploiting a narrow region of the solution space.

The algorithm behind Generalized Simulating Annealing is summarized in the following paragraphs based on \cite{tsallis1996generalized}. 
With a starting value of $x_1$, the size of the jump from $x_t$ to $x_{t+1}$ is determined based on a visiting distribution with a parameter $q_v$. The general form of the visiting distribution is as follows. 
\[
g_{q_v}(\Delta x_t) = c\frac{[T^V_{q_V}(t)]^d}{\{[T^V_{q_V}(t)]^e + (q_V - 1)b(
\Delta x_t)^2\}^{a/(q_v-1}}
\] where $a,b,c,d,e$ are constants that are calculated based on $q_V$ and the dimensions of the objective function and $T^V_{q_V}(t)$ is the visiting temperature at time t.
\[
T^V_{q_V}(t) = T_{q_V}(1)\frac{2^{q_V-1} -1}{(1+t)^{q_V-1}-1}
\]
The equation for the acceptance temperature is the same as above with $q_A$ used instead of $q_V$.
Based on the given objective function the objective value of the current state and the proposed state is calculated. If the proposed state has a lower objective function value than the current state then the proposed state is accepted. In the case it is not accepted, a random value between 0 and 1 is generated and a probability of acceptance is calculated as follows.
\[
P_{q_A}(x_t \rightarrow x_{t+1}) = \frac{1}{[1+(q_A - 1)(E(x_{t+1}) - E(x_t))/T^A_{q_A}]^{1/q_A-1}}
\]
The proposed state would then be accepted if the randomly generated probability is less than the acceptance probability. Finally, a new temperature value is calculated and the iterations continue forward until an optimum is discovered or the iterations have ended.

\subsubsection{Objective Function}
\label{sec:loss}
The loss function of the iterative refinement method can be summarized as follows.
\[
L(\left[\underline{X}\text{  }\underline{Y}\right]) = \frac{1}{k}\sum_{i=1}^k |s_i\Big(
\begin{bmatrix}
D_{i-1}\\
[\underline{X}\text{ }\underline{Y}]
\end{bmatrix}\Big)-m_{0i}|
\]
where $\mathbf{D_t} = [D_{t-1},\underline{X},\underline{Y}]$,$i \in \{Outlying, Skewed, \dots, Monotonic\}$ and $s_i(\big[\begin{smallmatrix}
D_{i-1}\\
[\underline{X}\text{ }\underline{Y}]
\end{smallmatrix}\big]),m_{0i}$ is the $i^{th}$ calculated and expected scagnostic measurement respectively.

\subsubsection{Algorithm}

\begin{algorithm}[H]
	\SetAlgoLined
	\KwResult{A set of 2D points $\mathbf{D}$}
	set number of points to add in an epoch as $n_0$, error variance as $\sigma^2$\;
	set $\underline{X},\underline{Y} \sim Uniform(0,1)$\;
	set $D_0 = [\underline{X}\text{ }\underline{Y}]$\;
	set starting values as $[\underline{X}\text{ }\underline{Y}]$\;
	\ForEach{$i^{th}$ epoch}{
			$\underline{X_{o}},\underline{Y_{o}} =\min\limits_{\underline{X_{p}},\underline{Y_{p}}} L([\underline{X_p}\text{ }\underline{Y_p}])$\;
			$D_i = \begin{bmatrix}
D_{i-1}\\
[\underline{X}\text{ }\underline{Y}]
\end{bmatrix}$\;
            $\underline{X_n} \sim \mathcal{N}(\underline{X_o},\sigma^2)$\;
            $\underline{Y_n} \sim \mathcal{N}(\underline{Y_o},\sigma^2)$\;
			set starting values as $[\underline{X_n}\text{ }\underline{Y_n}]$\;
	}
	return $\mathbf{D_N}$\;
	\caption{scatteR function}
\end{algorithm}

\subsubsection{Multiple Scagnostic measurement constraints}

In certain cases, we would want to mimic a given dataset to privatize the data and provide data that closely resembles the dataset with a wider variety. The scagnostics of the given dataset can be used as a target for scatteR to generate a sample that fits the scagnostics of the given dataset. This is also possible within scatteR as the algorithm itself uses a loss function that averages the absolute differences between the generated scagnostics and the expected scagnostic measurements over the number of measurement types. Hence, when multiple measurements are required to be satisfied, scatteR can provide a dataset that is as close as possible to the combination of expected scagnostic measurements.

\section{Experiments}
\label{sec:results}
\subsection{Experimental Setup}

The following experiments were performed on an Apple Macbook Pro with an M1 Chip and 16GB of RAM. R version 4.1.2 and OpenJDK version 11.0.12 was used while the results were visualized using the ggplot2 (\cite{ggplot2cite}) package from the tidyverse (\cite{tidyversecite}) ecosystem. The experiments were conducted with the global minimum target set to be at 0.0001 while the variance of the error term was set to 0.1. The hyperparameters of the simulated annealing algorithm were set to the defaults of the GenSA package (\cite{gensacite}) with the maximum number of iterations at 5000, initial temperature as 5230, the parameter of the visiting distribution as 2.62, and the parameter of the acceptance distribution as -5.0.

\subsection{Reliability of generated samples}

In order to assess the accuracy of this method in generating scatter plots that resemble the scagnostic measurements, an experiment was conducted where each measurement type and value was generated for multiple replicates. The resulting scatterplots were then fed into the scagnostic function to retrieve the generated measurements, which were then used to calculate the Root Mean Square Error (RMSE) 
as follows.
\[
RMSE(M,V) = \sqrt{\frac{\sum_{i=1}^{R}(s_M(\mathbf{D}) - V)^2}{R}}
\]
where $M \in \{\text{Outlying, Skewed,\dots, Monotonic}\}$ is the measurement type, $V \in \{0.0,0.5,1.0\}$ is the expected measurement value and $R = 20$ is the number of replicates performed for the experiment.

\begin{figure}[H]
\centering
\includegraphics[scale=1]{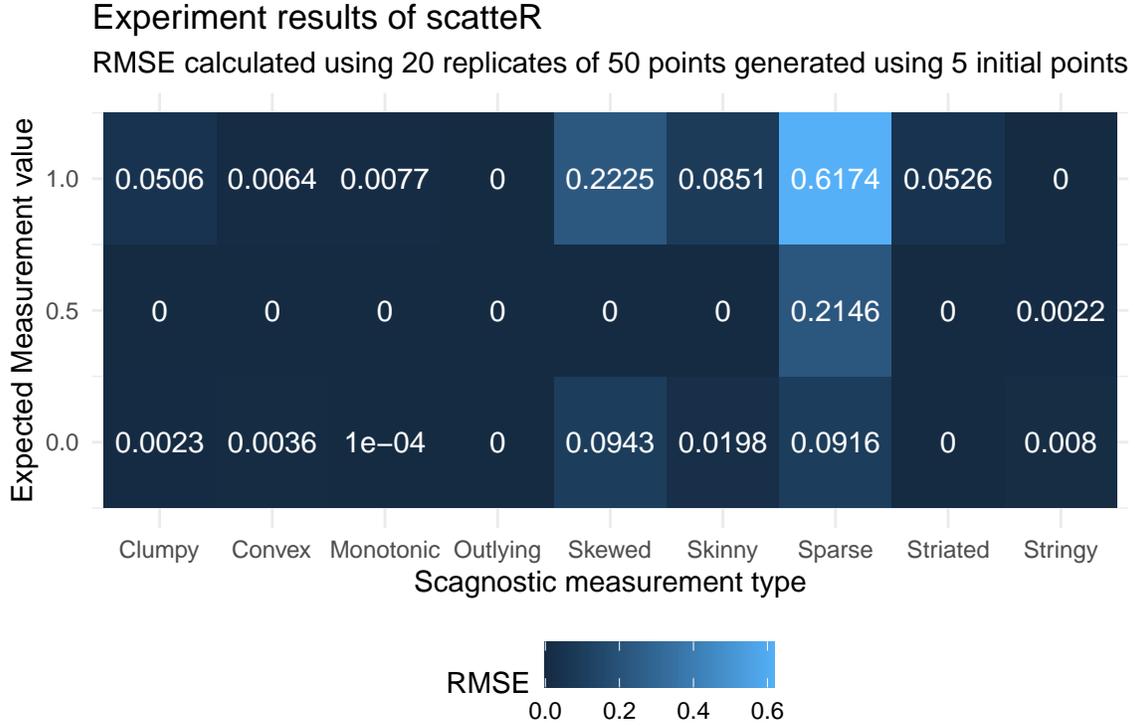}
\caption{Experiment results}
\label{fig:experimentresults}
\end{figure}

The results of the experiment are summarized in Figure \ref{fig:experimentresults} where scatteR has been capable of generating scatterplots that on average give scagnostic measurements which are approximately close to the expected scagnostic measurement.

\subsection{Computational time}

\begin{figure}[H]
    \centering
    \includegraphics[scale=0.75]{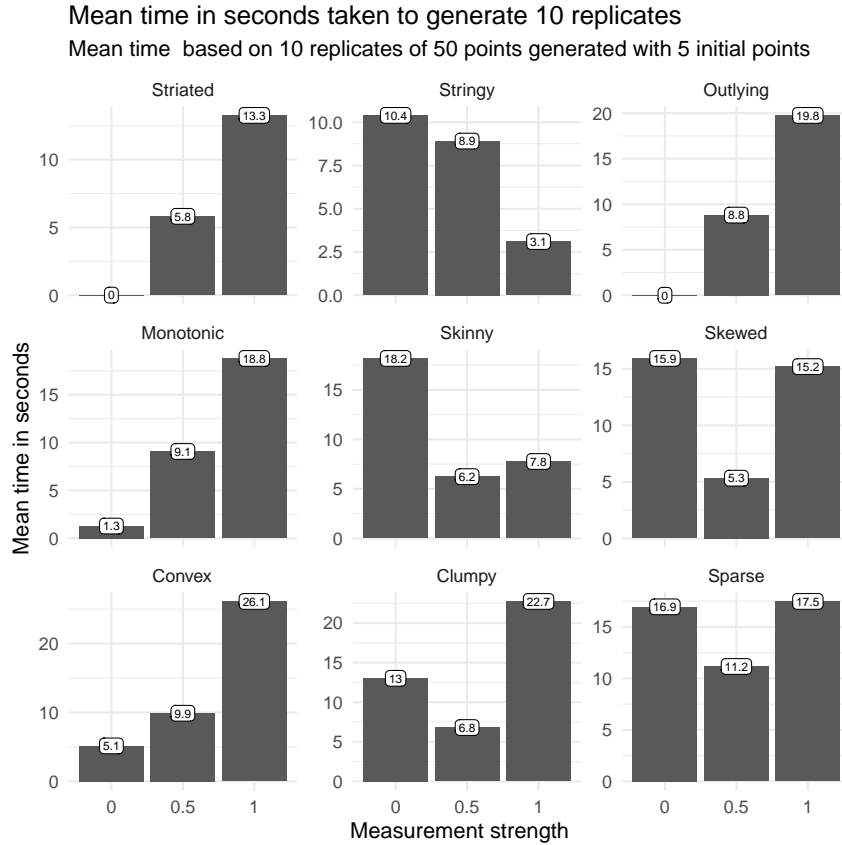}
    \caption{Computation time with respect to measurement type and strength}
    \label{fig:scag_type_value_comp_time}
\end{figure}

The scatteR algorithm contains an iterative refinement approach where simulated annealing is performed multiple times with increasing sample size and constant parameter size pertaining to the number of initial points. The computation time required to generate a fixed set of points should be studied to measure whether the scatteR algorithm is efficient in providing the samples required. The computational time of scatteR with respect to the different scagnostic measurements and their measurement strengths are presented in Figure \ref{fig:scag_type_value_comp_time}. 

All of the scatteR generations of 50 points using 5 initial points can be completed on average in under 30 seconds. In most cases, the scagnostics with lower strength can be generated quicker than the ones with higher strength. This could be due to the constraints imposed by the scagnostics algorithm, where a perfect 1.0 score in any measurement type would indicate a highly specific scatterplot arrangement. 

\begin{figure}[H]
    \centering
    \includegraphics[scale=0.14]{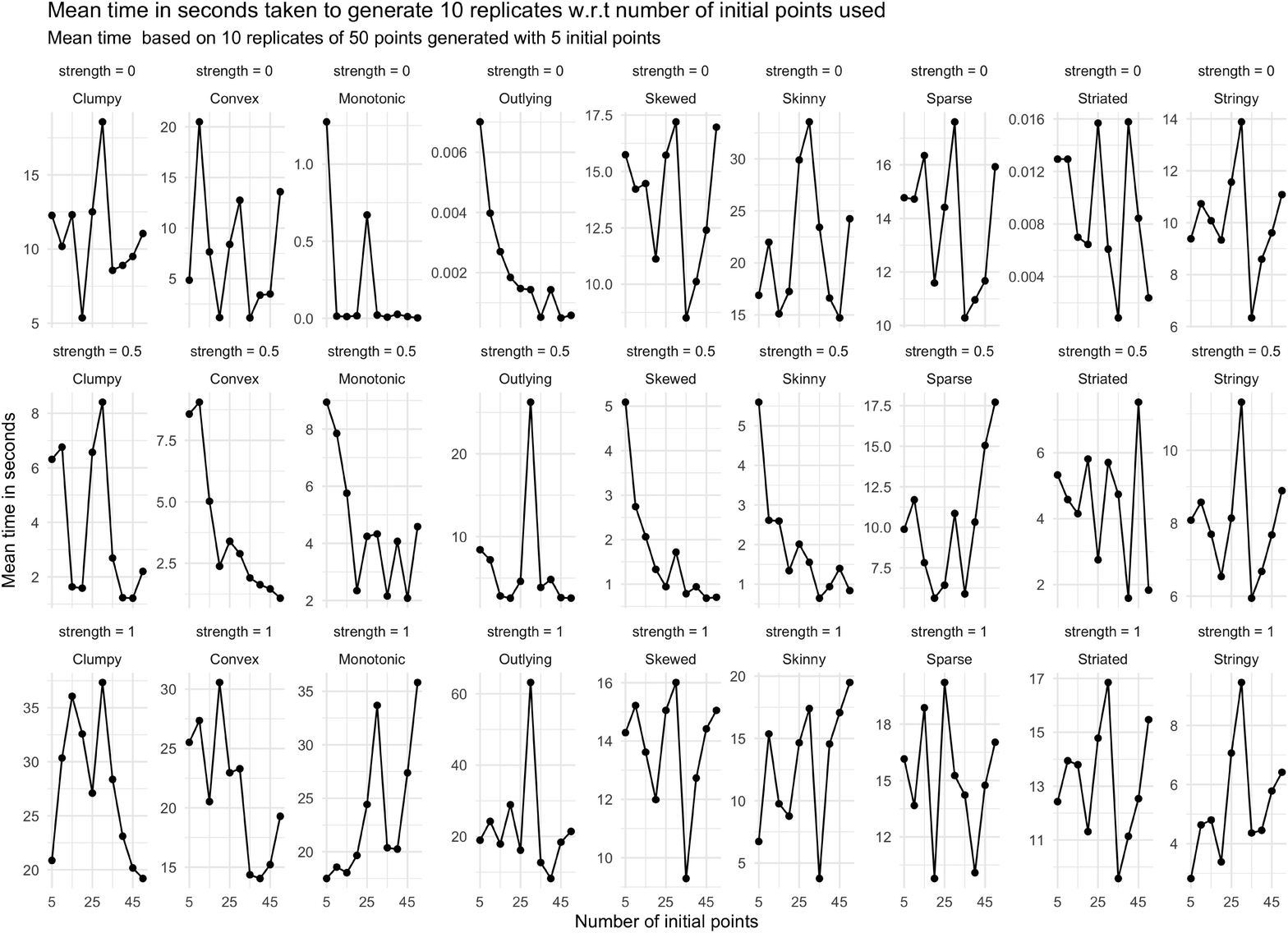}
    \caption{Computation time with respect to measurement type and strength and the number of initial points used}
    \label{fig:init_point_comp_time}
\end{figure}

From the available hyperparameters of the scatteR algorithm, the number of initial points has an impact as it is a deciding factor of the number of parameters that the simulated annealing algorithm has to optimize over in each iteration to make the final scatterplot resemble the expected scagnostics. Therefore, the time consumed by different configurations of the number of initial points was studied and the results are presented in Figure \ref{fig:init_point_comp_time}. 

As expected, a higher number of initial points does not increase the computation time. However, there is a spike of time consumed in the range of 25 to 40 initial points utilized.






\section{Conclusions}
\label{sec:conclusions}


In conclusion, the proposed method, scatteR, is a promising tool for generating scatterplots that resemble desired scagnostics measurements. Our results demonstrate that scatteR can find an optimal scatterplot by arranging a set of points iteratively using simulated annealing as the optimization algorithm, achieving a low RMSE value. Overall, scatteR offers a valuable contribution to the field of data generation and can aid in various statistical applications, such as data analysis, visualization and data augmentation. ScatteR can also be used to create datasets that mimic real-world scenarios, enabling researchers and practitioners to test and validate their models, algorithms, and hypotheses in a controlled setting.

In future work, we plan to enhance the methodology of scatteR by integrating the graph-theoretic features of the graphs used in scagnostic measurement calculations.  This extension will further strengthen the capability of scatteR to generate data that accurately captures the underlying patterns and anomalies within a collection of variables.

\bigskip
\begin{center}
{\large\bf SUPPLEMENTARY MATERIAL}
\end{center}

\begin{description}

\item[R-package for scatteR:] The R-package scatteR contains the code related to this paper which can be found at \url{https://github.com/janithwanni/scatteR}.

\item[Conflict of interest:] The authors declare that they have no conflicts of interest to disclose.

\end{description}
\bibliographystyle{jasa3}

\bibliography{Bibliography}
\end{document}